\documentstyle[12pt]{article}

\newcommand{\eq}{\begin{equation}}
\newcommand{\en}{\end{equation}}

\newcommand{\NP}[1]{Nucl.\ Phys.\ {\bf #1}}
\newcommand{\PL}[1]{Phys.\ Lett.\ {\bf #1}}

\newcommand{\CMP}[1]{Comm.\ Math.\ Phys.\ {\bf #1}}

\newcommand{\PRL}[1]{Phys.\ Rev.\ Lett.\ {\bf #1}}

\begin{document}

\hskip 10cm \vbox{\hbox{DFTT 21/92}\hbox{May 1992}}
\vskip 0.4cm
\centerline{\bf NEW METHODS OF INTEGRATION IN MATRIX MODELS }
\vskip 1.3cm
\centerline{A. D'Adda}
\vskip .6cm

\centerline{\sl Istituto Nazionale di Fisica Nucleare}
\centerline{\sl Sezione di Torino}
\centerline{\sl Dipartimento di Fisica
Teorica dell'Universit\`a di Torino}
\centerline{\sl via P.Giuria 1, I-10125 Turin,Italy}

\vskip 1.cm

\begin{abstract}
We discuss a new method of integration over matrix variables based on
a suitable  gauge choice in which the angular variables decouple from
 the eigenvalues at least for a class of two-matrix models. The
calculation of correlation functions involving angular variables is
simple in this gauge. Where the method is applicable it also gives an
extremely simple proof of the classical integration formula used to
reduce multi-matrix models to an integral over the eigenvalues.
\end{abstract}
\vfill
\hrule
\vskip.2cm
\noindent

\noindent

\hbox{\vbox{\hbox{$^{\diamond}${\it email address:}}\hbox{}}
 \vbox{\hbox{ Decnet=(39163::DADDA)}
\hbox{ Bitnet=(DADDA@TORINO.INFN.IT)}}}

\eject

\newpage
\setcounter{page}{1}
Matrix models have stimulated a great deal of interest in recent years
following the discovery \cite{DBG} that they provide a powerful tool in
studying two dimensional gravity and its coupling with $c \leq 1$
conformal matter beyond perturbation expansion in the genus \cite{rev}.
In spite of that there has been very small
progress in the actual mathematical tools used in solving matrix and
multi-matrix models.
In particular the multi-matrix models for which it is known how to
perform the integration over the angular variables are still the ones
listed in ref. ~\cite {Mehta}. In all these cases the integration
 over the angular variables is done by using one fundamental  formula:
\eq
\int d\mu(U)~e^{-cTr(U \Lambda^{(1)} U^{-1} \Lambda^{(2)})} = \left(
{ 2 \pi \over c} \right)^{N(N-1)/2} { 1 \over N!} { \det \left( e^{-c
\lambda_a^{(1)} \lambda_b^{(2)}} \right) \over
 \Delta(\Lambda^{(1)}) \Delta(\Lambda^{(2)})}
\label{1}
\en
where $U$ is a unitary $N \times N$ matrix,
$\Lambda^{(1)}$ and $\Lambda^{(2)}$ two diagonal matrices ($\Lambda_{ab}
^{(i)}=\delta_{ab} \lambda_a^{(i)}$) and
\eq
\Delta(\lambda^{(i)})=\prod_{1 \le b<a \le N}~(\lambda_a^{(i)} -
\lambda_b^{(i)})
\label{2}
\en
In eq. (\ref{1}) $d\mu (U)$ denotes the Haar measure of $SU(N)$ with the
following normalization:
\eq
\int d \mu (U) = { (2 \pi )^{N(N-1)/2} \over \prod_{p=1}^N p!}
\stackrel{\rm def}{=} \Omega_N
\label{1a}
\en
This equation was first derived, for a general compact Lie group,
 by Harish-Chandra ~\cite{Chandra} , rediscovered in the context of
matrix models in ~\cite{Itzykson} and fully exploited in ~\cite{Mehta}.
For a general proof of (\ref{1})  see also ~\cite{Alt}.

In this paper we will present an alternative way to perform the
integration over the angular variables, which consists in fixing the
gauge in such a way that the integration over the remaining variables is
trivial. When it is applicable (we shall discuss this point later)
this method offers the advantage that after the gauge fixing the angular
variables are decoupled from the eigenvalues, so that the calculation of
correlation functions involving angular variables  is greatly
simplified. Also when such decoupling does not occur, as in the quartic
couplings of the six vertex model (see page 26 in Ginsparg review paper
\cite{rev}), the dependence of
the action  on the angular variables is much simplified and it leaves
some hope that the integration over them might eventually be done.
Finally the method provides (in the cases discussed below) an extremely
simple algebraic proof of eq.(\ref{1}).

As a preliminary example let us consider the case of a one matrix model
 defined in terms of
a hermitian $N \times N$ matrix $A$ by the partition function
\eq
Z_N=\int dA e^{-Tr[V(A)]}
\label{3}
\en
The action $TrV(A)$ is invariant under the unitary transformation
\eq
A \to A^{'} = U^{-1} A U
\label{4}
\en
and so it depends only on the eigenvalues $\lambda_a$ of $A$.
We can treat the invariance under (\ref{4}) as an ordinary gauge
invariance and eliminate the unphysical degrees of freedom by a gauge
fixing procedure \footnote{A number of people must have been aware of
this procedure, to my knowledge it has appeared in the literature in
Ginsparg review paper \cite{rev} and in ~\cite{me}.}.
In the gauge where $A$ is diagonal
\eq
A_a^{~b} = \delta_a^{~b} \lambda_a
\label{5}
\en
an infinitesimal unitary transformation of parameters $\epsilon_a^{~b}$
 is given by
\eq
\delta A_a^{~b} = i[\epsilon,A]_a^{~b} = \epsilon_a^{~b} (\lambda_a -
\lambda_b)
\label{6}
\en
and the corresponding Faddeev-Popov determinant is
\eq
det  {\delta A_a^{~b} \over \delta \epsilon_c^{~d} } =
\prod_{a>b} (\lambda_a - \lambda_b)^2
\label{7}
\en
The partition function $Z_N$ is then reduced to an
integral over the eigenvalues:
\eq
Z_N = \Omega_N \int \prod_a d\lambda_a \prod_{a>b}
 (\lambda_a -\lambda_b)^2 e^{-\sum_a V(\lambda_a)}
\label{8}
\en
where $\Omega_N$ is the volume of the  gauge group  $SU(N)$ defined in
(\ref{1a}).

Let us consider next the case of a two matrix model defined
by the partition function:
\eq
Z_N^{(2)} = \int d\varphi d\varphi^{\dagger} e^{-Tr[V(\varphi)
 + \overline{V}(\varphi^{\dagger}) + c\varphi \varphi^{\dagger}]}
\label{9}
\en

where
 $\varphi$ and $\varphi^{\dagger}$ are respectively  a complex
 $N\times N$ matrix and its
hermitian conjugate. If one defines  $\varphi=A+iB$ this is clearly
equivalent to a model with two hermitian matrices $A$ and $B$.
The action in (\ref{9}) is invariant under unitary transformations of
$\varphi$:
\eq
\varphi \to \varphi^{'}=U^{\dagger}\varphi U~~~~~~~~,\varphi^{\dagger}
\to \varphi^{\dagger '}=U^{\dagger} \varphi^{\dagger} U
\label{10}
\en

It is always possible , by using the invariance (\ref{10}) , to reduce
$\varphi$ to a triangular form, that is:
\eq
\varphi_{ab} = 0~~~~a> b~~~~,\varphi^{\dagger}_{ab} = 0~~~~a<b
\label{11}
\en
Notice that in such gauge the diagonal elements of $\varphi$ and
$\varphi^{\dagger}$ coincide with the respective eigenvalues:
 $\varphi_{aa}
=\lambda_a$ and $ \varphi^{\dagger}_{aa} =\overline{ \lambda_a}$.
The advantage of the gauge fixing conditions (\ref{11}) is that the
corresponding Faddeev-Popov determinant is very simple and depends only
on the eigenvalues on $\varphi$ and $\varphi^{\dagger}$. In fact if one
denotes by $\Phi_{ab}$ the l.h.s. of the gauge conditions (\ref{11}),
that is
$\Phi_{ab}=\varphi_{ab}$ for $a>b$ and $\Phi_{ab} = \varphi^
{\dagger}_{ab}$ for $a<b$,
and by $\epsilon_{ab}$ the parameters of an infinitesimal unitary
transformation ($\epsilon_{ba} = \overline{\epsilon_{ab}}$), then it is
easy to show that:
\eq
det {\delta \Phi_{ab} \over \delta \epsilon_{cd}} = \Delta (\lambda)
\Delta (\overline{\lambda}) = \prod_{a>b} |\lambda_a - \lambda_b |^{2}
\label{12}
\en
With this choice of gauge the partition function (\ref{9}) takes the
form:
\begin{eqnarray}
Z_N^{(2)} &=& \Omega_N \int \ \prod_{a=1}^{N} d\lambda_a
 d\overline{\lambda_a} \prod_{a<b} d\varphi_{ab}
d\overline{\varphi}_{ab} \ \Delta(\lambda) \Delta(\overline{\lambda})
\nonumber \\ & &
 \exp \{ -\sum_{a=1}^{N} [V(\lambda_a) +
\overline{V}(\overline{\lambda_a}) + c \lambda_a \overline{\lambda_a}] -
c\sum_{a<b} \varphi_{ab} \overline{\varphi}_{ab} \}
\label{13}
\end{eqnarray}
As the angular variables appear quadratically in (\ref{13}) and are
decoupled from the eigenvalues , the corresponding gaussian integral is
trivial and gives:
\eq
Z_N^{(2)} = \left( {2 \pi \over c} \right)^{N(N-1)/2} \Omega_N
 \int \ \prod_{a=1}^{N} d\lambda_a
 d\overline{\lambda_a} \Delta(\lambda) \Delta(\overline{\lambda})
 e^{ -\sum_{a=1}^{N} [V(\lambda_a) +
\overline{V}(\overline{\lambda_a}) + c \lambda_a \overline{\lambda_a}]}
\label{14}
\en
 Eq.(\ref{14}) is the same as the the one considered for
instance in \cite{Mehta} , except that the two hermitian matrices
 are replaced here by complex conjugate matrices $\varphi$ and
$\varphi^{\dagger}$.
The novelty of our procedure is in the simple quadratic dependence of
the action from the remaining angular variables after fixing the gauge
as shown in eq.(\ref{13}) . This allows in particular to calculate
(or at least to reduce to integrals over the eigenvalues) correlation
functions of gauge invariant quantities , such as for instance
 $Tr\varphi^2 \varphi^{\dagger 2}$ and
$Tr \varphi \varphi^{\dagger} \varphi \varphi^{\dagger}$,
 that involve also angular variables.
In these cases the integration over $\varphi_{ab}$ and
 $\overline{\varphi}_{ab}$ can be done by simple Feynman diagrams
techniques.
These correlation functions  are the first terms of the
expansion in powers of the coupling constants of the partition function
generated by the action
\eq
S = Tr[V(\varphi)+ \overline{V}(\varphi^{\dagger})+c \varphi \varphi^
{\dagger} +g_1 \varphi \varphi \varphi^{\dagger} \varphi^{\dagger} +
g_2 \varphi \varphi^{\dagger} \varphi \varphi^{\dagger}
\label{15}
\en
which contains as a particular case the matrix model describing the
coupling of the six vertex model to gravity.
In (\ref{15}) the angular variables are not decoupled , in fact in the
 gauge (\ref{11}) we have for instance:
\begin{eqnarray}
 Tr   \varphi \varphi^{\dagger} \varphi \varphi^{\dagger}& =&  \sum
 \left\{ | \lambda_a |^4 + 2 | \lambda_a |^2 \left[ | \varphi_{ba} |^2 +
 | \varphi_{ab} |^2 \right] + 2  \lambda_a \overline{\varphi}_{ba}
\overline{\varphi}_{ac}  \varphi_{bc} \right. \nonumber \\&  &\left.
 + 2  \overline{\lambda}_a
\varphi_{ab} \varphi_{ca} \overline{\varphi}_{cb} +  \varphi_{ab}
\varphi_{cd} \overline{\varphi}_{cb} \overline{\varphi}_{ad} \right\}
\label{16}
\end{eqnarray}
where the $\varphi_{ab}$'s and the $\overline{\varphi}_{ab}$'s at the
 r.h.s. of the equation stand for the off-diagonal elements only (
$\varphi_{ab} = \overline{\varphi}_{ab} = 0$ for $a \geq b$).
In this situation the task of reducing the calculation of the partition
function to an integral over the eigenvalues is , beyond the first few
terms in the coupling constants expansion, a formidable one but not ,
according to some preliminary calculations, a completely hopeless
one.

It would obviously be very interesting to generalize these results to
the case of models of two hermitian matrices. However a hermitian matrix
cannot be cast in a triangular form and we could not find any other
gauge choice that offers the same advantages.
Still one can apply our method to a different model, which seems to be
closely related to the one of two hermitian matrices although we were
unable to prove a complete equivalence.
Let us consider two hermitian matrices $A_i$ ($i=1,2$) parameterized as
usual in the following way:
\eq
A_i = U_i \Lambda_i U_i^{\dagger}
\label{17}
\en
where $\Lambda_i$ are diagonal ($\Lambda_{i,ab}= \delta_{ab} \lambda_a^{
(i)}$) and $U_i$ are unitary matrices. Let us associate to any hermitian
matrix (\ref{17}) a real matrix $\hat{A}_i$ defined as follows:
\eq
\hat{A}_i = \hat{U}_i \Lambda_i \hat{U}_i^{-1}
\label{18}
\en
where $\hat{U}_i$ is the matrix of $SL(N,r)$ obtained from $U_i$ by a
"Wick rotation" of the group parameters corresponding to the real-
symmetric generators of $SU(N)$. More explicitly, if we denote by $T^{(
ab)}$ the antisymmetric generators of $SU(N)$ and by $S^{(ab)}$ the
symmetric ones, we have:
\eq
U_i = e^{i\tau_{ab}^{(i)}T^{(ab)} + i \sigma_{ab}^{(i)} S^{ab}} \;
\stackrel{\sigma_{ab} \rightarrow i \sigma_{ab}}{\longrightarrow} \;
\hat{U}_i = e^{i\tau_{ab}^{(i)}T^{(ab)} - \sigma_{ab}^{(i)} S^{ab}}
\label{19}
\en
It should be noticed that the $\hat{A}^{(i)}$'s are not generic real
matrices since their eigenvalues are constrained to be real.
By means of a $SL(N,r)$ transformation the two matrices $\hat{A}_{ab}^{(
i)}$ can be cast into a triangular form, more precisely:
\eq
\hat{A}_{ab}^{(1)} = 0~~~~~a> b~~~~~~~~~\hat{A}_{ab}^{(2)}
 = 0~~~~~a< b
\label{20}
\en
As before the diagonal elements coincide in this gauge with the
eigenvalues.
Consider now the partition function
\eq
\hat{Z}_N^{(2)} = \int_{\cal D} d\hat{A}_{ab}^{(1)} d\hat{A}_{ab}^{(2)}
e^{-Tr \left[ V_1(\hat{A}^{(1)}) + V_2 (\hat{A}^{(2)}) +ic\hat{A}^{(1)}
\hat{A}^{(2)} \right] }
\label{21}
\en
where the integration is restricted to a domain $\cal D$ corresponding
to real eigenvalues for the real matrices $\hat{A}^{(1)}$ and $\hat{A}^{
(2)}$. The action is invariant under transformations of the non compact
group $SL(N,r)$ ; by fixing the gauge according to (\ref{20}) and by
calculating the Faddeev-Popov determinant in complete analogy to eq.
(\ref{12}) one finds:
\begin{eqnarray}
\hat{Z}_N^{(2)} & = & \hat{\Omega}_N \int \prod_{a=1}^N d\lambda_a^{(1)}
d\lambda_a^{(2)} \prod_{a<b} d\hat{A}_{ab}^{(1)} d \hat{A}_{ba}^{(2)}
\Delta (\lambda^{(1)}) \Delta (\lambda^{(2)}) \cdot \nonumber \\
& & \exp  \left\{ - \sum_{a=1}^N \left[ V_1 (\lambda_a^{(1)}) + V_2 (
\lambda_a^{(2)} ) + ic\lambda_a^{(1)} \lambda_a^{(2)} \right] - ic
\sum_{a<b} \hat{A}_{ab}^{(1)} \hat{A}_{ba}^{(2)} \right\}
\label{22}
\end{eqnarray}
where $\hat{\Omega}_N$ is the infinite volume of the gauge group. Notice
that the coupling constant $c$ in front of the term $Tr \hat{A}^{(1)}
\hat{A}^{(2)}$ is multiplied by $i$ in (\ref{22}). Such "Wick rotation"
is needed in order to have a finite result from the integration over the
remaining angular variables in (\ref{22}) which would otherwise be
divergent due to the non compact nature of the group.
The final result is:
\begin{eqnarray}
\hat{Z}_N^{(2)}& =& \hat{\Omega}_N \left( {2 \pi \over c} \right)^
{N(N-1)/2} \int \prod_{a=1}^N d\lambda_a^{(1)} d\lambda_a^{(2)}
 \Delta(\lambda^{(1)}) \Delta(\lambda^{(2)}) \cdot \nonumber \\
 & &  e^{-\sum_a \left[ V_1(\lambda_a^{(1)})+ V_2(\lambda_a^{(2)})
 + ic \lambda_a^{(1)}\lambda_a^{(2)} \right] }
\label{23}
\end{eqnarray}
This result coincide (after substituting $ic$ back with $c$) with the
standard result obtained in \cite{Mehta} for hermitian matrices. However
it should be noticed that , as in the case of complex conjugate matrices
, eq. (\ref{22}) gives some extra information, namely that the angular
variables are decoupled from the eigenvalues and appear quadratically in
the action.
All the considerations following eq. (\ref{13}) and (\ref{14}) can be
applied to the present case by simply replacing $\varphi$ and $\varphi^
{\dagger}$ with $\hat{A}^{(1)}$ and $\hat{A}^{(2)}$, in particular eqs.
similar to (\ref{15}) and (\ref{16}) can be easily derived.

It is possible also to prove the $SL(N,r)$ analogue of eq.(\ref{1}),
which is in fact a simple consequence of (\ref{22}) and of
\eq
d\hat{A}^{(i)} = \prod_a d\lambda_a^{(i)} \Delta^2(\lambda^{(i)}) d \mu(
\hat{U}_i)
\label{24}
\en
By inserting (\ref{18}) and (\ref{24}) into eq. (\ref{21}) one finds:
\begin{eqnarray}
\hat{Z}_N & = &\hat{\Omega}_N \int \prod_a d\lambda_a^{(1)}d
 \lambda_a^{(2)} d\mu(\hat{U}) \Delta^2(\lambda^{(1)})
 \Delta^2(\lambda^{(2)}) \cdot \nonumber \\ & & e^{-
\sum_a \left[ V_1(\lambda_a^{(1)}) + V_2(\lambda_a^{(2)}) \right] -icTr
\left( \hat{U} \Lambda^{(1)} \hat{U}^{-1} \Lambda^{(2)} \right) }
\label{25}
\end{eqnarray}
It is now sufficient to compare eqs (\ref{25}) and (\ref{22}) and to
take into account that the integral over the eigenvalues is weighted by
arbitrary functions which are symmetric under separate permutations of
the eigenvalues of $\hat{A}^{(1)}$ and $\hat{A}^{(2)}$. The result is:
\begin{eqnarray}
\lefteqn{N! \int  d\mu(\hat{U}) e^{-ic Tr( \hat{U} \Lambda^{(1)}
 \hat{U}^{-1} \Lambda^{(2)})}  = } \nonumber \\ & &= \sum_P { (-1)^{|P|}
 e^{-ic \sum_a (P\lambda^{(1)} )_a \lambda_a^{(2)}}  \over
 \Delta(\lambda^{(1)}) \Delta (\lambda^{(2)})} \cdot \int \prod_{a<b} d
 \hat{A}_{ab}^{(1)} d \hat{A}_{ba}^{(2)} e^{-ic \sum_{a<b}
 \hat{A}_{ab}^{(1)} \hat{A}_{ba}^{(2)} } \nonumber \\ & &= \left( {2 \pi
\over c} \right)^{N(N-1)/2} {\det \left( e^{-ic \lambda_a^{(1)}
\lambda_b^{(2)}} \right) \over  \Delta(\lambda^{(1)}) \Delta
 (\lambda^{(2)})}
\label{26}
\end{eqnarray}
where the sum is over all permutations $P$ on the eigenvalues $\lambda_a
^{(1)}$.
 In deriving eq.(\ref{26}) one has used the fact that the l.h.s. is
invariant under permutations of the eigenvalues. In fact it is easy to
show that for any permutation $P$ it exists a matrix $\hat{U}_P \in SL(N
,r)$ such that
\eq
P\Lambda^{(1)} = \hat{U}_P \Lambda^{(1)} \hat{U}_P^{-1}
\label{27}
\en
Eq. (\ref{26}) can be generalized by including in the original partition
function (\ref{21}) higher order interactions involving $\hat{A}^{(1)}$
and $\hat{A}^{(2)}$. For instance the addition of a quartic term $ig Tr
\hat{A}^{(1)} \hat{A}^{(2)} \hat{A}^{(1)} \hat{A}^{(2)}$ to  the action
in (\ref{21}) would result in an extra term
\eq
-ig \left( Tr \hat{U} \Lambda^{(1)} \hat{U}^{-1} \Lambda^{(2)} \hat{U}
 \Lambda^{(1)} \hat{U}^{-1} \Lambda^{(2)} \right)
\en
in the exponential at the l.h.s. of (\ref{26}). Correspondingly the
exponential at the l.h.s. would acquire a contribution of the type
\begin{eqnarray}
 & &-ig  \left\{ \sum \right. \left[ \lambda_a^{(1)~2}
 \lambda_a^{(2)~2} \right. + 2 \lambda_a^{(1)} \lambda_a^{(2)}
 \left( \hat{A}_{ab}^{(1)} \hat{A}_{ba}^{(2)} + \hat{A}_{ab}^{(2)}
 \hat{A}_{ba}^{(1)} \right) +2 \lambda_a^{(1)} \hat{A}_{ab}^{(2)}
 \hat{A}_{bc}^{(1)} \hat{A}_{ca}^{(2)} +  \nonumber \\
 & &+2 \left. \left.\lambda_a^{(2)} \hat{A}_{ab}^{(1)} \hat{A}_{bc}^{(2)}
 \hat{A}_{ca}^{(1)} + \hat{A}_{ab}^{(1)}
\hat{A}_{bc}^{(2)} \hat{A}_{cd}^{(1)} \hat{A}_{da}^{(2)} \right]
\right\}
\end{eqnarray}
where $\hat{A}^{(i)}_{ab}$ denote the off-diagonal elements only and the
  sum is then restricted to $\hat{A}^{(1)}_{ab}=0$ for $a>b$ and $\hat{A}
^{(2)}_{ab} =0$ for $a<b$.
Clearly in this case the integration over $\hat{A}_{ab}^{(1)}$ and
$\hat{A}_{ab}^{(2)}$ at the r.h.s. of eq. (\ref{26}) involves the
eigenvalues and it cannot be done explicitly, at least so far.

It should be noticed that our proof of eq.(\ref{26}) and its
 generalizations  follows in
a purely algebraic way from comparing two different choices of gauge in
the partition function.
\vskip 2 truecm
\bf Acknowledgments \rm : It is a pleasure to acknowledge precious
daily discussions with M. Caselle .

\end{document}